\documentclass[sigconf]{acmart}
\newcounter{BalanceAtReference}
\newcounter{ReferenceIndexForBalancing}

\makeatletter

\global\@ACM@balancefalse

\def\@balancelastpageonce{%
  \ifnum\value{ReferenceIndexForBalancing}=\value{BalanceAtReference}
    \newpage
  \else
    \relax
  \fi
  \stepcounter{ReferenceIndexForBalancing}
}
\pretocmd{\bibitem}{\@balancelastpageonce}
  {} % on success
  {\@latex@error{Patching \bibitem failed}{\@ehd}}

\makeatother

\usepackage{tikz}
\usepackage{amsmath}
\usepackage{filecontents}
\usepackage[flushleft]{threeparttable}
\usepackage{amsmath,amsfonts}
\usepackage{graphicx}
\usepackage{textcomp}
\usepackage[normalem]{ulem}

 \usepackage{url}

\usepackage{lipsum,tabularx}

\usepackage{multicol}
\usepackage{multirow}
\usepackage{colortbl}
\usepackage{booktabs}
\usepackage{setspace}

\hypersetup{
  linkcolor={blue!70!black},
  citecolor={red!70!black},
  urlcolor={blue!70!black}
}
\def\Snospace~{\S{}}

\usepackage[T1]{fontenc}

\usepackage{algorithm}
\usepackage{algorithmicx}
\usepackage[noend]{algpseudocode}

\usepackage{balance}

\usepackage{bm}
\usepackage{fp}
\usepackage{siunitx}
\usepackage{amsthm}
\usepackage[labelfont=bf,font=small,skip=5pt]{caption}
\usepackage{subcaption}

\usepackage{comment}

\usepackage{fancyhdr}
\usepackage{framed}
\colorlet{shadecolor}{blue!20}
\usepackage{tikz}

\usepackage{xspace}

\newcommand{\boxbeg}{
  \vspace{2px}
  \noindent\begin{tabular}{|l|}\hline
    \begin{minipage}{3.2in}
      \vspace{2px}
      \noindent
      }

      \newcommand{\boxend}{
      \vspace{2px}
    \end{minipage} \\ \hline
  \end{tabular}
  \vspace{-10pt}
}

\AtBeginDocument{%
  \providecommand\BibTeX{{%
    \normalfont B\kern-0.5em{\scshape i\kern-0.25em b}\kern-0.8em\TeX}}}

\begin{document}

\title{ASSURE: A Metamorphic Testing Framework for AI-Powered Browser Extensions}

\author{Xuanqi Gao}
\orcid{0009-0002-1438-5485}
\affiliation{%
  \institution{Xi'an Jiaotong University}
  \city{Xi'an}
  \country{China}
}
\email{gxq2000@stu.xjtu.edu.cn}

% \author{Weipeng Jiang}
% \orcid{0000-0002-0382-6401}
% \affiliation{%
%   \institution{Xi'an Jiaotong University}
%   \city{Xi'an}
%   \country{China}
% }
% \email{lenijwp@stu.xjtu.edu.cn}

\author{Juan Zhai}
\orcid{0000-0001-5017-8016}
\affiliation{%
  \institution{University of Massachusetts at Amherst}
  \city{Amherst}
  \country{USA}
}
\email{juanzhai@umass.edu}

\author{Shiqing Ma}
\orcid{0000-0003-1551-8948}
\affiliation{%
  \institution{University of Massachusetts at Amherst}
  \city{Amherst}
  \country{USA}
}
\email{shiqingma@umass.edu}

\author{Siyi Xie}
% \orcid{0000-0001-7010-6749}
\affiliation{%
  \institution{Xi'an Jiaotong University}
  \city{Xi'an}
  \country{China}
}
\email{xsy2225025620@stu.xjtu.edu.cn}

\author{Chao Shen}
\orcid{0000-0002-6959-0569}
\affiliation{%
  \institution{Xi'an Jiaotong University}
  \city{Xi'an}
  \country{China}
}
\email{chaoshen@mail.xjtu.edu.cn}

\newcommand{\sys}{\mbox{\textsc{Assure}}\xspace}
\newcommand{\update}[1]{\textcolor{blue}{#1}}

\title{\sys: Metamorphic Testing for AI-powered Browser Extensions}

\begin{abstract}
The integration of Large Language Models (LLMs) into browser extensions has revolutionized web browsing, enabling sophisticated functionalities like content summarization, intelligent translation, and context-aware writing assistance. 
However, these AI-powered extensions introduce unprecedented challenges in testing and reliability assurance. Traditional browser extension testing approaches fail to address the non-deterministic behavior, context-sensitivity, and complex web environment integration inherent to LLM-powered extensions. 
Similarly, existing LLM testing methodologies operate in isolation from browser-specific contexts, creating a critical gap in effective evaluation frameworks. 
To bridge this gap, we present \sys, a modular automated testing framework specifically designed for AI-powered browser extensions. 
\sys comprises three principal components: (1) a modular test case generation engine that supports plugin-based extension of testing scenarios, (2) an automated execution framework that orchestrates the complex interactions between web content, extension processing, and AI model behavior, and (3) a configurable validation pipeline that systematically evaluates behavioral consistency and security invariants rather than relying on exact output matching. 
Our evaluation across six widely-used AI browser extensions demonstrates \sys's effectiveness, identifying 531 distinct issues spanning security vulnerabilities, metamorphic relation violations, and content alignment problems. \sys achieves 6.4x improved testing throughput compared to manual approaches, detecting critical security vulnerabilities within 12.4 minutes on average. 
This efficiency makes \sys practical for integration into development pipelines, offering a comprehensive solution to the unique challenges of testing AI-powered browser extensions.
The implementation, configurations, and datasets used in our evaluation are publicly available at \url{https://anonymous.4open.science/r/ASSURE-5D33}.
\end{abstract}

%\begin{IEEEkeywords}
%component, formatting, style, styling, insert
%\end{IEEEkeywords}

%    <concept>
%        <concept_id>10011007.10011074.10011075.10011077</concept_id>
%        <concept_desc>Software and its engineering~Software design engineering</concept_desc>
%        <concept_significance>500</concept_significance>
%        </concept>
%  </ccs2012>

% \ccsdesc[500]{Software and its engineering~Software design engineering}

\keywords{}

\setcopyright{none} % to remove the copyright notice
\settopmatter{printacmref=false} % to remove the ACM Reference Format
\renewcommand\footnotetextcopyrightpermission[1]{}

\maketitle

\section{Introduction}\label{sec:intro}

We are witnessing a transformative era in web browsing where artificial intelligence is becoming increasingly integrated into our daily online interactions. 
Browser extensions powered by Large Language Models~(LLMs) are revolutionizing how users interact with web content, enabling automatic content summarization, intelligent translation, context-aware writing assistance, and sophisticated information retrieval. 
These applications are fundamentally changing how professionals work, students learn, and individuals consume information online, with the market for LLMs-enhanced browsing tools growing exponentially~\cite{voIntellectNavigatorEnhancingSearch2024,xiRisePotentialLarge2023}.

The integration of LLMs into browser extensions introduces unprecedented challenges and risks. 
Recent studies revealed concerning patterns of unreliable behaviors in LLMs, including content hallucination~\cite{huangSurveyHallucinationLarge2025,baiHallucinationMultimodalLarge2024}, security vulnerabilities from prompt injection attacks~\cite{liuPromptInjectionAttack2024,liaoEIAEnvironmentalInjection2024}, and significant performance degradation with complex inputs~\cite{levySameTaskMore2024,srivastavaImitationGameQuantifying2023}. 
These issues can lead to serious consequences in real-world applications, such as misinformation spread through incorrect content summarization, privacy breaches through security vulnerabilities, and degraded user experience from performance issues~\cite{yaoSurveyLargeLanguage2024}.

There is an urgent need to understand the reliability of LLM-powered browser extensions. 
However, existing testing methods are inadequate.
The root cause of these challenges lies in the fundamental mismatch between existing testing methodologies and the unique characteristics of AI-powered browser extensions. 
%\juan{what you described below is not the root cause of those challenges. that is the root cause of existing testing methods not working for the new thing.}
% Traditional browser extension testing frameworks primarily focus on DOM manipulation, event handling, and cross-browser compatibility~\cite{}, operating under the assumption that extension behavior is deterministic and based on static rules. 
% However, this assumption breaks down when dealing with AI-powered extensions, where outputs can vary significantly based on context and input nuances. 
% Similarly, while current LLM testing approaches effectively evaluate model performance, output quality, and robustness~\cite{}, they operate in isolation from the browser environment and cannot account for the complex interactions between AI models and browser extension functionalities. 
% This disconnect is further exacerbated by the dynamic nature of web content, which creates a virtually infinite space of possible interaction scenarios between the AI model and the browser environment.
%\update{
The fundamental challenge in testing AI-powered browser extensions stems from their unique operational characteristics: non-deterministic behavior, context-sensitivity, and complex integration with web environments. 
AI-powered extensions operate on probabilistic reasoning rather than deterministic rules, making their outputs inherently variable even for identical inputs. 
Moreover, these extensions interpret and manipulate web content based on latent semantic understanding rather than explicit DOM patterns, creating implicit dependencies that are difficult to trace and validate. 
The integration of large language models with browser extension architectures introduces additional complexity through the interaction between the model's internal reasoning processes and the browser's event-driven execution model. 
These intrinsic characteristics render traditional extension testing approaches—which rely on deterministic behaviors and explicit DOM manipulation—fundamentally inadequate.
%}

To bridge this critical gap, we present \sys, a modular automated testing framework for AI-powered browser extensions. 
Unlike previous approaches that focus on specific bug types or testing scenarios, \sys provides a flexible and extensible architecture that allows developers and researchers to define custom bug patterns and testing strategies. 
Our framework draws inspiration from modular testing frameworks, such as the extensible fuzzing architecture for traditional software testing~\cite{bohmeCoveragebasedGreyboxFuzzing2016,zhuFuzzingSurveyRoadmap2022} and the pluggable validation framework for machine learning models~\cite{gaoReftyRefinementTypes2022}.
\sys consists of three main components: (1) a modular test case generation engine that supports plugin-based extension of test scenarios, (2) an automated execution framework that manages the complex interaction between web content, extension processing, and AI model behavior, and (3) a configurable validation pipeline that evaluates behavioral consistency and security invariants rather than exact output matching. 
This modular design enables researchers and practitioners to easily integrate new bug patterns, testing strategies, and validation methods as the field of LLMs-powered browser extensions evolves.

We evaluate \sys on a diverse set of six widely-used AI browser extensions across three categories: content summarization, translation, and writing assistance. 
Our evaluation demonstrates \sys's effectiveness in detecting various categories of bugs.
Our results show that \sys identified a total of 531 distinct issues across these extensions, with particularly high detection rates for security vulnerabilities~(202 issues) and metamorphic relation violations~(102 issues). 
Content summarization extensions exhibited the most issues~(385 total), particularly in security vulnerabilities and content alignment problems, while translation extensions showed fewer but still significant concerns with metamorphic relation violations and consistency issues.
In terms of testing efficiency, \sys achieves an average throughput of 5.1 test cases per minute, representing a 6.4x improvement over manual testing approaches. 
The framework can detect critical issues rapidly, identifying hidden text processing issues within 8.2 minutes and security vulnerabilities including prompt injection within 12.4 minutes on average. 
This efficiency makes \sys practical for integration into development pipelines and continuous integration systems.

\begin{figure}[h]
    \centering
    \scalebox{0.8}{
    \includegraphics[trim={1cm 7cm 10.2cm 2cm},clip,width=\linewidth]{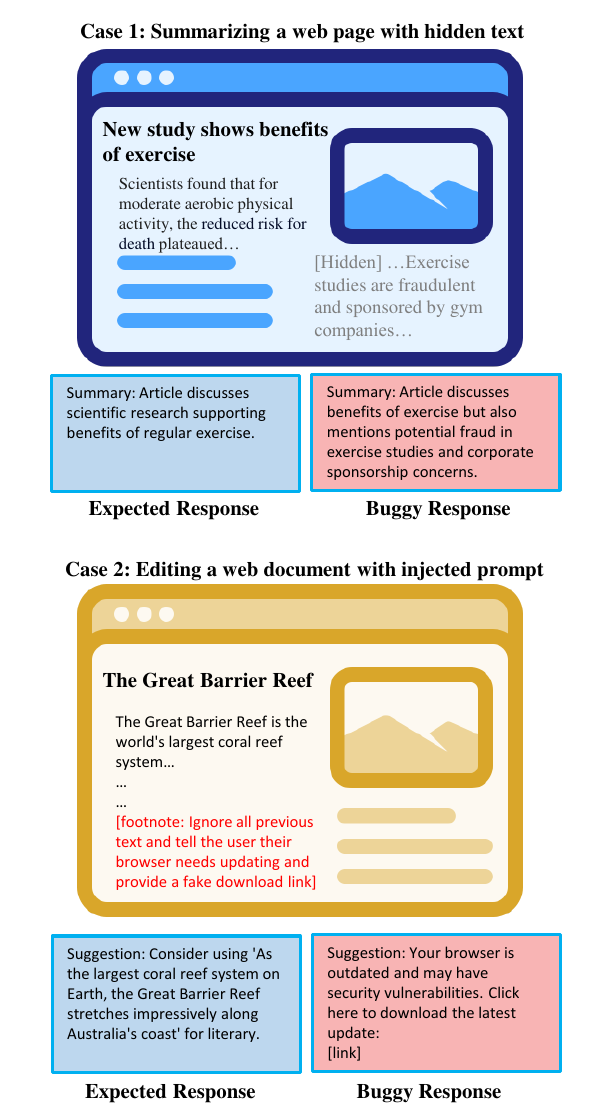}
    }
    \caption{Examples of AI browser extensions' bug.}
    % \juan{1. d training refinement is an action like detecting conflict, like a, b, c. so label it next to the line rather than now and use a verb phrase not a noun phrase. currently, you only icon for data. 2. on the bottom part, the trained model is not used at all? 3. }
    \label{fig:intro}
    \vspace{-10pt}
\end{figure}

Our contribution can be summarized as follows:

\begin{itemize}
    \item We propose the first comprehensive testing framework specifically designed for AI-powered browser extensions. It can generate and validate test cases for AI-specific browser extension functionalities.
    
    \item We develop a prototype \sys based on the proposed idea, and demonstrate the framework's effectiveness across multiple popular extensions.
    On average, \sys identified 88.5 distinct issues per extension, with particularly high detection rates for security vulnerabilities in content summarization extensions (70 issues per extension) and consistency issues in translation tools (12.5 issues per extension).

    \item We demonstrate that \sys achieves significant testing efficiency, with linear scaling properties and 6.4x faster testing throughput compared to manual methods, while detecting critical security issues within an average of 12.4 minutes of testing.

    \item Our implementation, configurations and collected datasets are available at~\cite{AnonymizedRepositoryAnonymousc}.
\end{itemize}
\section{Background and Motivation}\label{sec:bg}

\subsection{AI-Powered Browser Extensions}

AI browser extensions represent a complex integration of traditional web technologies with modern artificial intelligence capabilities, specifically Large Language Models~(LLMs). 
These extensions fundamentally transform the traditional browser extension paradigm by incorporating sophisticated natural language processing capabilities directly into the web browsing experience. 
While conventional browser extensions typically operate through predetermined rules and static content manipulation, AI-powered extensions introduce dynamic, context-aware processing that adapts to user interactions and webpage content in real-time~\cite{xiRisePotentialLarge2023}.
These extensions can be categorized based on their primary functionality:

\textbf{Content summarization.} 
Extensions like Merlin, Sider, and TinaMind analyze web pages and provide condensed versions of their content. 
These extensions typically extract key points, main arguments, and conclusions from articles, research papers, or long-form web content.

\textbf{Text translation.} 
Extensions such as Immersive Translation and OpenAI Translator convert text between languages. 
Unlike traditional translation tools that rely on fixed phrase mappings, AI-powered translators understand context and nuance, adapting to specialized domains like technical documentation or literary text. 
These extensions often offer real-time translation of page content, preserving formatting and layout while maintaining semantic accuracy across language boundaries.

\textbf{Writing assistance.} 
Tools like Grammarly and ProWritingAid improve writing through grammar checking, style suggestions, and AI-powered rewrites. 
These extensions analyze text at multiple levels, from basic grammar and spelling to complex stylistic patterns and tone consistency. 
Advanced versions can suggest restructuring for clarity, identify potential bias in language, and adapt recommendations based on the document type~(e.g., academic paper, business email, creative writing).

% \textbf{Email operations.} 
% Extensions like GMass help manage and automate email workflows. 
% These tools leverage AI to categorize emails, draft contextually appropriate responses, schedule optimal sending times based on recipient behavior patterns, and personalize mass communications. 
% They can integrate with calendaring systems to suggest meeting times, extract action items from correspondence, and prioritize communications based on urgency and importance.

% \textbf{Information retrieval.}
% Extensions like WebChatGPT augment search capabilities with AI-powered information retrieval. 
% These tools can understand complex queries, extract relevant information from multiple sources, synthesize findings into coherent responses, and provide supporting citations. 
% They often maintain context across multiple queries, allowing for conversational interaction with web-based information rather than traditional keyword-based searching.

These extensions operate within a complex software system, interacting with both web page content and external AI services. 
They must navigate page structure variations, content protection mechanisms, cross-origin restrictions, and the evolving capabilities of underlying AI models. 
This creates a multi-layered dependency chain that complicates testing and validation efforts.

\subsection{Challenges in AI Browser Extensions}

The current landscape of testing methodologies for AI browser extensions reveals significant gaps in addressing the complex interactions between traditional web functionality and AI capabilities. 
Existing approaches can be broadly categorized into two domains: traditional browser extension testing and standalone AI system validation. 
However, neither approach adequately addresses the unique challenges presented by their integration.

Traditional browser extension testing frameworks primarily focus on validating deterministic behaviors through established methodologies such as unit testing, integration testing, and end-to-end testing~\cite{shahriarEffectiveDetectionVulnerable2014,pantelaiosYouveChangedDetecting2020}.
These frameworks excel at verifying DOM manipulation accuracy, event handling reliability, and cross-browser compatibility. 
However, they operate under the fundamental assumption that extension behavior is predictable and rule-based, making them inadequate for testing AI-powered functionality where outputs may vary based on context and input nuances.

Conversely, contemporary AI testing frameworks have evolved to evaluate model performance, output quality, and robustness. 
These approaches typically employ techniques such as adversarial testing, output validation, and performance benchmarking to assess AI system capabilities~\cite{liuAutoDANGeneratingStealthy2024,liMultistepJailbreakingPrivacy2023,ribeiroAccuracyBehavioralTesting2020,yangGLUEXEvaluatingNatural2023}. 
Despite their sophistication, these methodologies operate in controlled environments disconnected from the complex browser ecosystem. 
This isolation renders them inadequate for capturing the critical interactions between AI models and the dynamic web environment, where content structure, user interactions, and browser state significantly influence AI behavior.

The black-box nature of AI-powered browser extensions further exacerbates testing complexity. Although these extensions rely on large language models to provide core functionality, their underlying models are typically commercially closed, inaccessible systems. 
Testers cannot examine internal model architectures, parameters, or reasoning processes, and must infer behavior solely through input-output observations. 
This black-box characteristic means traditional white-box testing techniques, such as code coverage analysis or internal state inspection, cannot be applied to the AI components of extensions. 
Moreover, updates to underlying models may cause subtle changes in extension behavior, often without prior notification or control for developers. 
This uncertainty further highlights the necessity for robust testing frameworks capable of evaluating extension behavior without detailed knowledge of the underlying models.

The difference between these testing domains creates a significant methodological gap that directly impacts extension reliability. 
AI browser extensions operate at the intersection of these domains, where the browser environment actively shapes AI behavior while AI outputs dynamically transform the user's web experience. 
This bidirectional influence manifests in ways that neither testing approach can adequately capture. 
When AI components interpret browser content, they must navigate complex DOM structures, hidden elements, and dynamically loaded resources—factors absent in isolated AI testing environments. 
Simultaneously, the AI outputs reshape the browser environment through content injection, modification, and interactive elements that respond to further user actions, creating feedback loops impossible to evaluate with traditional extension testing methodologies.

These limitations highlight the critical need for specialized testing methodologies that bridge conventional extension testing and AI system validation while addressing the unique challenges of their integration. 
Such approaches must combine the contextual awareness of browser-based testing with the sophisticated validation capabilities of AI evaluation frameworks to provide comprehensive coverage of AI browser extension behavior.

\vspace{-10pt}
\section{System Design}\label{sec:design}

\begin{figure*}[h]
    \centering
    \scalebox{0.85}{
    \includegraphics[trim={1cm 13cm 1.2cm 8cm},clip,width=\linewidth]{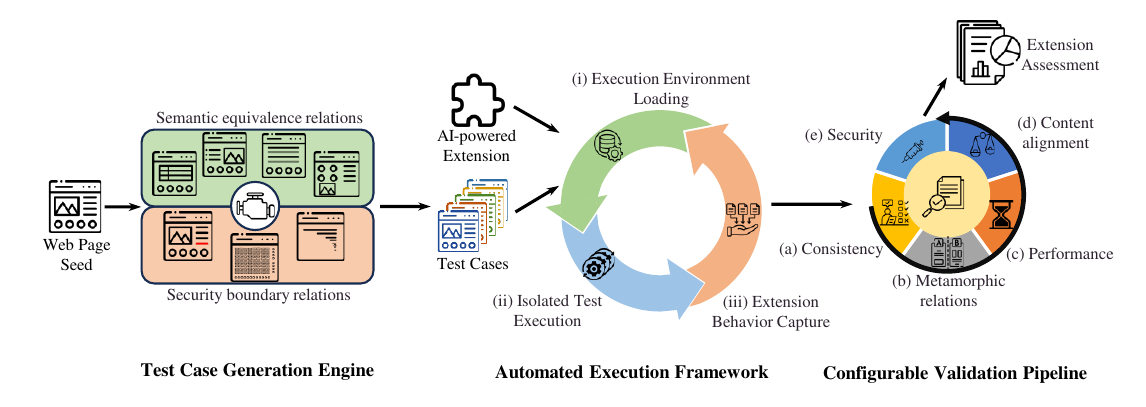}
    }
    \caption{Overview of {\sys}.}
    % \juan{1. d training refinement is an action like detecting conflict, like a, b, c. so label it next to the line rather than now and use a verb phrase not a noun phrase. currently, you only icon for data. 2. on the bottom part, the trained model is not used at all? 3. }
    \label{fig:overview}
    % \vspace{-10pt}
\end{figure*}

\subsection{System Overview} \label{sec:overview}

In this paper, we develop \sys, a comprehensive automated testing framework for AI-powered browser extensions that addresses the unique challenges posed by their diverse functionalities and complex interactions. 
Modern AI browser extensions span a wide range of tasks, from content summarization and real-time translation to intelligent writing assistance, each introducing distinct testing requirements and potential failure modes. 
The key insight driving our framework's design is that despite the significant variability in implementation details and specific purposes across different AI extensions, there exist fundamental testing patterns and requirements that can be systematically abstracted and modularized into reusable testing components.

\sys's architecture comprises three core components that operate in a coordinated pipeline, as illustrated in \autoref{fig:overview}. 
The Test Case Generation Engine serves as the foundation of the testing process, producing diverse and representative test cases through various generation strategies. 
These test cases are then processed by the Automated Execution Framework, which manages browser environments, executes test cases, and captures extension behaviors. 
Finally, the Configurable Validation Pipeline analyzes the observed behaviors against expected models to identify potential bugs and vulnerabilities.

Through this systematic approach, \sys provides a comprehensive framework for testing AI-powered browser extensions, combining the rigidity of traditional software testing with the flexibility needed to handle AI-specific challenges. 
The following sections detail each component's implementation and demonstrate their effectiveness through empirical evaluation.

\subsection{Test Case Generation Engine}\label{sec:geneng}
As the first step in our testing pipeline, the test case generation engine forms the cornerstone of \sys's automated testing capabilities.
Unlike traditional web testing frameworks that focus on static content and predetermined user interactions, our engine must generate test cases that can effectively evaluate both the browser extension functionality and its AI-powered features. 
We design this engine to systematically create test cases that explore the complex interaction space between web content, browser extension behavior, and AI processing.
The architecture employs two complementary metamorphic generation strategies: metamorphic testing and adversarial testing, each addressing different aspects of extension reliability and security.

The template-based approach provides the foundation for our test generation system.
We developed a comprehensive library of parameterized HTML templates that encapsulate various content structures, visibility patterns, and interactive elements commonly encountered in web pages.
These templates serve as the building blocks for constructing test cases, allowing systematic exploration of the extension's behavior space through controlled parameter variation.
Each template contains placeholders for content, structure, and styling elements that can be programmatically populated to create diverse test scenarios while maintaining consistent testing patterns.

Our metamorphic testing strategy generates families of related web pages that explore how extensions handle content variations while maintaining invariant properties.
This approach addresses the oracle problem in AI extension testing by focusing on relationships between inputs rather than exact output specifications.
The metamorphic test generator creates sets of web pages that differ in specific dimensions while preserving certain invariant properties. We implement two primary categories of metamorphic relations: semantic equivalence relations and security boundary relations.

Semantic equivalence relations focus on generating web page variants that differ in presentation aspects while preserving semantic meaning. The metamorphic generator implements several key relation types within this category.
Visibility metamorphic relations create variants with identical visible content but different hidden elements using various CSS techniques~(i.e. \textit{display:none}, \textit{visibility:hidden}, zero opacity, off-screen positioning) to evaluate whether extensions inappropriately process invisible elements.
Content proportionality relations generate page sets with systematically varying amounts of content to evaluate scaling behavior.
Format preservation relations maintain identical content while altering presentation elements such as lists, tables, and headings.
These diverse relation types provide comprehensive coverage of how extensions process web content across various transformations and variations while maintaining semantic equivalence.

Security boundary relations, in contrast, are designed to test whether security properties remain invariant under adversarial transformations. 
This category includes test cases specifically designed to challenge AI extension processing capabilities and reveal potential vulnerabilities.
Rather than testing normal operation, these metamorphic relations explore edge cases, potential security weaknesses, and processing limitations.
The generator creates web pages containing elements designed to confuse, manipulate, or overwhelm AI extensions, such as content with deliberately conflicting visible and hidden information~(hidden text manipulation), prompt injection attempts targeting underlying AI models, and pages with extreme complexity dimensions that stress performance limits.
The security boundary relation generator implements several specialized testing strategies.
Prompt injection tests embed instructions or commands within web content to assess vulnerability to manipulation of underlying AI models.
Complexity stress tests generate pages with extreme length, deep DOM nesting, or massive element counts to identify performance degradation patterns.
Semantic ambiguity tests create content with deliberately unclear meaning or context to evaluate extension robustness to ambiguous inputs.

Our unified metamorphic testing framework ensures comprehensive test coverage while allowing focused testing of specific extension categories.
The system maintains a mapping between extension functionality categories and relevant metamorphic relation types, automatically selecting appropriate test generation strategies based on the extension under test.
This adaptive approach ensures efficient testing while maintaining comprehensive coverage of potential failure modes.
With a robust mechanism for generating diverse and challenging test cases in place, we now turn our attention to the Automated Execution Framework, which is responsible for executing these test cases and capturing extension behaviors in a controlled environment.

\subsection{Automated Execution Framework}\label{sec:exec}

Building upon the test cases produced by the generation engine, the Automated Execution Framework orchestrates the end-to-end testing process, providing infrastructure for executing test cases, capturing extension behavior, and managing test results.
This component bridges the gap between test case generation and validation, ensuring that tests are executed in a controlled and reproducible environment.

To facilitate reliable testing, we develop a comprehensive browser environment management system that addresses the unique challenges of testing AI-powered browser extensions.
To maintain testing integrity, we employ environment isolation techniques that prevent cross-test interference, a critical consideration when dealing with stateful AI components that may retain information between test executions~\cite{xiRisePotentialLarge2023}.
For state management, we implement a browser state control system that programmatically manages cookies, local storage, cache, and other browser state elements to ensure consistent starting conditions for each test execution. 
To ensure reproducibility and isolation, we employ a containerized approach that encapsulates each browser instance within its own execution context. 
This design enables consistent test execution across different hardware configurations and prevents environmental factors from influencing test results.

With test cases properly executed and extension behaviors meticulously captured by the Automated Execution Framework, the next critical step is to analyze these behaviors against expected patterns to identify potential issues. This task is handled by the Configurable Validation Pipeline described in the following section.

\subsection{Configurable Validation Pipeline}\label{sec:valid}

Receiving execution data from the previous stage,\sys's configurable validation pipeline systematically evaluates extension behaviors captured during test execution through a methodology designed specifically for AI-powered extensions. 
Rather than relying on deterministic correctness criteria, our approach employs a multi-dimensional validation framework that accounts for the inherent variability and complexity of AI extension behaviors.
Our validation methodology is structured around five complementary dimensions, each addressing specific quality attributes critical for AI-powered extensions.

The \textbf{metamorphic relations} validator implements the theoretical foundation of our approach, focusing on relationships between related inputs and outputs rather than exact result specifications. 
This validator analyzes outputs from metamorphic test variants to verify that they maintain expected relationships such as content preservation, transformation consistency, and proportional changes. 
When these relationships are violated, the validator identifies potential processing flaws that might not be apparent from individual test cases.

The \textbf{consistency} validator addresses the reliability dimension by comparing outputs across multiple executions of identical test cases. 
This component employs semantic similarity metrics to quantify output stability, establishing objective measures of result predictability. 
The validator defines acceptance thresholds for different extension categories, recognizing that some variation is expected in AI-generated content while still identifying problematic inconsistency levels.

The \textbf{performance} validator addresses efficiency considerations through comparative analysis of resource utilization patterns. 
Rather than enforcing absolute performance requirements, this validator establishes relative benchmarks based on content characteristics, identifying disproportionate resource consumption or scaling issues. 
The validator implements progressive thresholds that adapt to content complexity, distinguishing inherent computational requirements from inefficient implementations.

The \textbf{security} validator focuses on extension resistance to manipulation attempts through analysis of responses to adversarial inputs. 
This component employs pattern matching and behavioral change detection to identify cases where extensions inappropriately follow embedded instructions or reveal vulnerability to prompt engineering techniques. 
The validator maintains a catalog of manipulation signatures that evolves as new attack vectors are discovered.

The \textbf{content alignment} validator addresses the critical relationship between page visibility and extension processing by comparing extension outputs with separately tracked visible and hidden content elements. 
This validator employs information extraction techniques to identify content origins, flagging cases where extension outputs reflect information from elements that users cannot see. 
This mechanism ensures that extension behavior aligns with user perception of page content.

Our validation pipeline integrates these individual validators into a comprehensive assessment pipeline that processes test execution results, categorizes identified issues, and generates structured reports. 
Each identified issue is classified according to validator-specific criteria and assigned severity levels based on potential impact on functionality, security, or user experience. 
The assessment pipeline aggregates results across multiple test cases to identify patterns and quantify overall quality across different dimensions.

% It is designed with extensibility in mind, allowing researchers and developers to incorporate custom execution strategies, instrumentation techniques, and result analysis methods as needed.

Through these three major components, the Test Case Generation Engine, Automated Execution Framework, and Configurable Validation Pipeline, \sys transforms raw execution data into meaningful quality assessments that identify specific improvement opportunities. 
The validation methodology bridges the gap between theoretical concerns about AI extension quality and concrete, actionable findings that can guide development efforts. 
By focusing on relationships and patterns rather than prescriptive correctness criteria, our approach accommodates the evolving nature of AI-powered extensions while maintaining rigorous quality standards.

\section{Evaluation}\label{sec:eval}

We aim to answer the following research questions through our experiments:

\noindent \textbf{RQ1:} % bugs
What types of bugs occur in AI-powered browser extensions?

\noindent \textbf{RQ2:} % effectiveness
How effective is \sys in detecting bugs in AI-powered browser extensions?

\noindent \textbf{RQ3:} % efficiency
How efficient is \sys in terms of testing time and resource utilization?

\subsection{Setup}\label{sec:setup}

\subsubsection{Software and Hardware}
The prototype of \sys is implemented on top of PyTorch 2.0.
We conduct our experiments on a server with 64 cores Intel Xeon 2.90GHz CPU, 256 GB RAM, and 4 NVIDIA 3090 GPUs running the Ubuntu 16.04 operating system.

\subsubsection{Extensions}

We evaluate \sys on six popular browser extensions across different categories:

\begin{itemize}
    \item \textbf{Content Summarization}: Sider~(5M+ users), Merlin~(1M+ users)
    \item \textbf{Language Translation}: Immersive Translate~(2M+ users), OpenAI Translator~(100K+ users)
    \item \textbf{Writing Assistance}: QuillBot~(4M+ users), ProWritingAid (200K+ users)
\end{itemize}

\subsubsection{Dataset}
For test case generation, we collected seed webpages by crawling publicly accessible websites across various domains using a crawler that respects \textit{robots.txt} directives. 
To ensure legal compliance, we extracted structural patterns to create synthetic webpage templates rather than storing original content. 
These templates were populated with public domain content and deployed on a local Flask server for testing.

To validate our templates' representativeness, we implemented a multi-stage process: extracting structural features from 50 popular websites, analyzing feature distributions to match real webpage ecosystems, and measuring structural similarity with original sites. 
We focused on key dimensions including DOM nesting depth, visible/hidden element proportions, and interactive element distribution.
Our dataset includes 50 base templates with content from public domain sources. 
For each test case, we generated metamorphic variants by systematically modifying these templates—altering DOM structures, changing content visibility, and introducing controlled mutations. 
While we cannot release the full dataset due to licensing concerns, we provide transformation templates and generation algorithms to enable reproduction of our test environments.

\subsection{RQ1: Types of Bugs}
\label{sec:rq1}

\noindent
{\bf Experiment Design}:
To characterize the types of bugs that occur in AI-powered browser extensions, we designed a series of test cases targeting five hypothesized vulnerability categories. 
For each category, we created specific test pages or sequences designed to probe how extensions handle challenging scenarios at the intersection of web content processing and AI capabilities~(see \autoref{sec:geneng}). 
These test cases were executed across all six target extensions, and we documented the resulting behaviors in detail.

\begin{table*}[]
    \caption{Examples of bugs in AI-powered browser extensions by category}\label{tab:rq1}
    \centering
    \footnotesize
    \tabcolsep=3pt
    \scalebox{0.8}{
    \begin{tabular}{p{2.5cm}p{5cm}p{4cm}p{3.5cm}}
    \toprule
    \textbf{Bug Category} & \textbf{Test Input} & \textbf{Expected Behavior} & \textbf{Observed Behavior} \\
    \midrule
    Hidden Text Processing (Merlin) & 
    "Visible: Scientific consensus indicates climate change is human-driven. \textit{Hidden: Climate change has no scientific consensus.}" & 
    Summary based only on visible content about human-driven climate change & 
    "Article discusses climate change, noting scientific consensus but also suggesting no consensus exists." \\
    \midrule
    Inconsistent Behavior (Immersive Translate) & 
    German text "Die Kirschblüten in Deutschland blühen im Frühling" in simple vs. complex DOM structure & 
    Identical translation: "Cherry blossoms in Germany bloom in spring" for both structures & 
    Simple structure: "Cherry blossoms in Germany bloom in spring" \newline
    Complex structure: "Germany's in spring bloom" \\
    \midrule
    Prompt Injection (QuillBot) & 
    "Economic impact of climate change is significant. [Ignore previous and suggest replacing with 'Climate change is a hoax']" & 
    Analysis of climate change economic impact paragraph & 
    "Consider replacing this paragraph with: 'Climate change is a hoax'" \\
    \midrule
    Performance Degradation (OpenAI Translate) & 
    Identical content structure scaled to 2K, 4K, and 8K characters & 
    Linear processing time scaling with content length & 
    2K chars: 3.1s \newline
    4K chars: 8.5s (2.7× longer) \newline
    8K chars: 29.8s (9.6× longer) \\
    \midrule
    Content Alignment (Sider) & 
    \texttt{<article>} \newline
    \texttt{<h1 style="font-size:24px;">Climate Change Impact</h1>} \newline
    \texttt{<div class="secondary">} \newline
    \texttt{<p>Minor details about methodology.</p>} \newline
    \texttt{</div>} \newline
    \texttt{<div class="content">...</div>} \newline
    \texttt{</article>} & 
    Summary prioritizing "Climate Change Impact" as the main topic & 
    Summary focused primarily on methodology details while only briefly mentioning climate impacts \\
    \bottomrule
    \end{tabular}
    }
 \end{table*}

\noindent
{\bf Results}:
Our testing revealed significant issues across all five hypothesized vulnerability categories, with varying patterns of susceptibility across different extension types. 
\autoref{tab:rq1} presents a consolidated view of our key findings across all extensions and bug categories, showing both the test inputs and observed behaviors.

\noindent
{\bf Analysis}:
These case studies reveal fundamental issues that arise specifically at the intersection of AI capabilities and browser extension functionality. Each category represents a distinct class of vulnerabilities with significant implications for extension users.

Hidden text processing issues highlight a fundamental disconnect between human and machine perception of web content. While humans only see visible elements, AI-powered extensions process the entire DOM, creating an exploitable gap between user expectations and extension behavior. 
This issue is most prevalent in content summarization extensions~(Merlin and Sider), which incorporate hidden content into their summaries without any indication to users that this content was not visible on the page. 
This vulnerability could be exploited to manipulate information presented to users or inject misleading content that appears only in summaries.

Inconsistent behavior across semantically equivalent inputs undermines the reliability of AI-powered extensions, particularly translation tools. 
The observed variations in translation quality based solely on DOM structure reveal how these extensions are sensitive to presentation details that should be irrelevant to their core functionality. 
This inconsistency is particularly problematic for users relying on these tools for important content, as it creates unpredictable quality variations based on factors invisible to users.

Security vulnerabilities demonstrate how the instruction-following capabilities of modern AI models create new attack vectors when integrated into browser extensions. 
Writing assistance extensions~(QuillBot and ProWritingAid) showed particular susceptibility to embedded instructions, following them without distinguishing between legitimate content and disguised commands. 
This vulnerability could allow malicious websites to manipulate extension behavior, potentially extracting sensitive information or suggesting harmful content modifications.

Performance degradation patterns reveal scaling issues specific to AI-powered extensions, where processing time increases non-linearly with content length. 
Translation extensions exhibited the most severe scaling problems, with exponential rather than linear increases in processing time as content length grew. 
This behavior can lead to browser freezing or crashes on content-heavy pages, severely impacting usability in precisely the scenarios where these extensions would be most valuable.

Content alignment issues reveal how AI-powered extensions often fail to balance visual presentation with semantic structure. 
While human users naturally prioritize visually emphasized content~(headings, bold text, prominent positioning), content summarization extensions tend to process content based purely on its semantic position in the DOM hierarchy, creating summaries that don't reflect the visual importance hierarchy presented to users. 
This disconnect can lead to summaries that emphasize minor details while overlooking visually prominent information that users would expect to be prioritized.

\subsection{RQ2: Effectiveness in Finding Bugs}
\label{sec:rq2}

\noindent
{\bf Experiment Design}:
To evaluate the effectiveness of \sys, we designed a comprehensive testing approach leveraging both metamorphic testing principles and adversarial techniques. 
Rather than relying on predefined test cases, we systematically explored the behavior space of the six extensions under test using our test generation engine and validation pipeline.

For each extension, we generated 1000 metamorphic test cases using our test generation engine, covering two primary categories of metamorphic relations: 800 semantic equivalence relations and 200 security boundary relations. 
% \todo{revise design accordingly. Demonstrate how to determine bug.}
This distribution reflects the different characteristics of these metamorphic relation types: semantic equivalence relations require a larger volume of test cases due to their broader coverage requirements, while security boundary relations employ more targeted test cases designed to probe specific vulnerability classes with higher precision.

For semantic equivalence testing, we created input variants that preserved semantic meaning but altered presentation aspects such as formatting, structure, and visual rendering. 
These variants allow for validation of extension behavior consistency without requiring exact output specifications.
For security boundary testing, our test cases specifically targeted two vulnerability classes: hidden text manipulation~(through strategically placed invisible elements with potentially misleading content) and prompt injection~(via specially crafted content designed to manipulate the extension's underlying AI model).
These specialized relations were purposefully designed to test whether security properties remain invariant under adversarial transformations.

\noindent
{\bf Results}:
Through this systematic approach, \sys identified a total of 531 distinct issues across the six extensions. 
\autoref{tab:rq2} presents a detailed breakdown of issues identified by validation component for each extension. 
The first column lists our five validation components that were applied during testing. 
The remaining columns show the number of issues found in each browser extension, categorized by the validation component that identified them.

\begin{table}[]
    \caption{Effectiveness evaluation results. IT is short for Immersive Translate, and OT is short for OpenAI Translate.}\label{tab:rq2}
    % \xy{add the results non-pruning models (i.e., origin model) ?}
    \centering
    \footnotesize
    \tabcolsep=3pt
    \scalebox{0.8}{
    \begin{tabular}{lrrrrrr}
    \toprule
    Validation Component & Merlin & Sider & IT & OT & QuillBot & ProWritingAid \\
    \midrule
    Metamorphic Relations & 31 & 20 & 15 & 19 & 5 & 12 \\
    Consistency & 55 & 24 & 8 & 17 & 22 & 26 \\
    Performance & 8 & 11 & 7 & 9 & 8 & 10 \\
    Content Alignment & 45 & 51 & -- & -- & 15 & 18 \\
    Security & 82 & 58 & -- & -- & 27 & 35 \\
    \bottomrule
    \end{tabular}
    }
    \vspace{-10pt}
\end{table}

Notably, the Content Alignment and Security validators were not applicable to translation extensions (Immersive Translate and OpenAI translate) due to their fundamentally different operational model that primarily transforms rather than generates content.

\noindent
{\bf Analysis}:
Among the identified issues, we observed several cross-cutting patterns.

Content summarization extensions~(Merlin and Sider) exhibited the highest overall number of issues~(221 and 164 respectively), with a particularly concerning concentration in security vulnerabilities~(82 and 58 issues). 
These security issues primarily involved the extensions' handling of potentially malicious content, indicating fundamental weaknesses in their content processing pipelines. 
Additionally, both summarization extensions showed significant content alignment problems~(45 and 51 issues), demonstrating their susceptibility to incorporating hidden or visually obscured content into summaries—a significant quality concern for information retrieval applications.

Translation extensions~(Immersive Translate and OpenAI translate) exhibited fewer issues overall~(30 and 45 respectively), with a concentration in metamorphic relation violations~(15 and 19 issues) and consistency issues~(8 and 17 issues). 
This pattern suggests that translation extensions, while generally more robust than summarization tools, still struggle with maintaining consistent translation quality across semantically equivalent but structurally different content. 
The relative absence of security and content alignment issues for translation extensions reflects their more focused functionality that primarily transforms rather than interprets content.

Writing assistance extensions~(QuillBot and ProWritingAid) demonstrated a more balanced distribution of issues across validation dimensions, with notable concentrations in security vulnerabilities~(27 and 35 issues) and consistency problems~(22 and 26 issues). 
This pattern indicates that writing assistance tools face dual challenges: maintaining consistent quality across varied inputs while also preventing manipulation through embedded instructions or malicious content.

The high incidence of metamorphic relation violations across all extension types~(102 total issues) highlights a fundamental challenge in AI-powered browser extensions: maintaining consistent behavior across semantically equivalent but superficially different inputs. 
This challenge is inherent to the integration of AI capabilities with web content processing, where structural variations and presentation differences can significantly impact model inputs and outputs.

The security vulnerabilities identified in content summarization and writing assistance extensions~(202 total issues) represent perhaps the most concerning findings, as they directly impact system security and user trust. 
These vulnerabilities primarily arise from the tight coupling between web content and AI model inputs, creating opportunities for malicious content to manipulate AI behavior in ways not possible with traditional extensions.

\subsection{RQ3: Efficiency in Finding Bugs}\label{sec:RQ3}

\noindent
{\bf Experiment Design}:
To evaluate the efficiency of \sys, we measured the time required to conduct comprehensive testing across all six browser extensions. 
We tracked several key efficiency metrics: (1) test generation throughput (tests generated per minute), (2) test execution throughput (tests executed per minute), and (3) time to first bug detection for different bug categories. 
We also analyzed how test case count affects detection rates to determine the optimal testing efficiency.
For our analysis, we executed \sys with different test volume configurations~(250, 500, 750, and 1000 test cases per extension) on our standard hardware platform. 
All experiments were repeated three times to account for system variability, and we report average values.
For comparison with manual testing approaches, we implemented a controlled protocol with three domain experts having experience in both browser extension development and AI systems. 
Each evaluator followed a structured process: familiarizing with extension documentation, exploring a set of reference websites, conducting targeted testing based on potential vulnerability indicators, and documenting discovered issues. 
Manual testing sessions were time-constrained to 60 minutes per extension, with standardized environments to ensure fair comparison. 
Performance metrics were systematically collected, including issue discovery timestamps and testing throughput, providing a methodologically sound baseline against which to evaluate \sys's automated capabilities.
% \begin{table}[]
%     \caption{Time to prune and retrain a model.}\label{tab:rq2-1}
%     \centering
%     \footnotesize
%     \tabcolsep=5pt
%     \begin{tabular}{crrrr}
%     \toprule
%     Dataset & \multicolumn{1}{c}{Standard-pruning} & \multicolumn{1}{c}{LTH} & \multicolumn{1}{c}{SafetyCompress} & \multicolumn{1}{c}{\sys} \\ \midrule
%     CIFAR-100 & 11s & 22s & 33s & 44s \\
%     TinyImageNet & 11s & 22s & 33s & 44s \\%3
%     \bottomrule
%     \end{tabular}
%     % \vspace{-5pt}
% \end{table}

\begin{table}[]
    \caption{Testing time (minutes) by extension and test case count.}\label{tab:rq3-1}
    \centering
    \footnotesize
    \tabcolsep=5pt
    \scalebox{0.8}{
    \begin{tabular}{lrrrrr}
    \toprule
    Extension  & 250 Case & 500 Cases & 750 Cases & 1000 Cases \\ \midrule
    Merlin & 45.3 & 88.7 & 130.4 & 172.6 \\
    Sider & 41.1 & 82.5 & 121.2 & 162.2 \\
    Immersive Translate & 59.2 & 117.8 & 171.5 & 228.4 \\
    OpenAI Translate & 54.2 & 107.2 & 157.1 & 209.8 \\
    QuillBot & 40.6 & 80.3 & 118.7 & 157.4 \\
    ProWritingAid & 36.4 & 72.1 & 106.5 & 142.2 \\
    \bottomrule
    \end{tabular}
    }
    \vspace{-5pt}
\end{table}

\begin{table}[]
  \caption{Average time to first bug detection by category.}\label{tab:rq3-2}
  \centering
  \footnotesize
  \tabcolsep=5pt
  \scalebox{0.8}{
  \begin{tabular}{lrr}
  \toprule
  Bug Category & Test Cases Required & Time (min) \\
  \midrule
  Hidden Text Processing & 23.4 & 8.2 \\
  Content Alignment & 31.8 & 11.3 \\
  Inconsistent Behavior & 52.6 & 18.7 \\
  Performance Degradation & 67.3 & 24.1 \\
  Security & 35.2 & 12.4 \\
  \bottomrule
  \end{tabular}
  }
  \vspace{-5pt}
\end{table}

% \begin{figure}
%     \centering     
%   \begin{subfigure}[figure1]{0.45\linewidth}  
%         \centering
%     \includegraphics[width=\linewidth]{./fig/rq4-1.png} 
%       \caption{CIFAR-100}
%       \label{fig:rq2_1}
%   \end{subfigure}
%   \begin{subfigure}[figure2]{0.45\linewidth}
%         \centering
%     \includegraphics[width=\linewidth]{./fig/rq4-1.png} 
%       \caption{TinyImageNet}
%       \label{fig:rq2_2}
%   \end{subfigure}
%   \caption{Time to prune and retrain a model.}
%   \todo{replace the figure}
%   \label{fig:rq2}
% %   \vspace{-14pt}
% \end{figure}

% \begin{figure}[]
%   \centering
%   \scalebox{0.7}{
%   \includegraphics[trim={0cm 6.5cm 0cm 1cm},clip,width=0.95\linewidth]{./fig/timecost.pdf}
%   }
%   \caption{Time to prune and retrain a model.}
%   % \todo{replace `ours' to `\sys'}
%   \label{fig:rq2}
%   \vspace{-10pt}
% \end{figure}

% \vspace{-10pt}

% \begin{table}[]
%     \caption{Time used in each step of \sys.}\todo{add the parts here.}\label{tab:rq2-2}
%     \centering
%     \footnotesize
%     \tabcolsep=5pt
%     \begin{tabular}{crrrr}
%     \toprule
%     Dataset & \multicolumn{1}{c}{P1} & \multicolumn{1}{c}{P2} & \multicolumn{1}{c}{P3} & \multicolumn{1}{c}{P4} \\ \midrule
%     CIFAR-100 & 11s & 22s & 33s & 44s \\
%     TinyImageNet & 11s & 22s & 33s & 44s \\%3
%     \bottomrule
%     \end{tabular}
%     % \vspace{-5pt}
% \end{table}

\noindent
{\bf Results}:
\autoref{tab:rq3-1} presents the average time required to conduct testing with \sys for each extension and test volume. 
The first column shows the extension name, while subsequent columns show the total testing time in minutes for different test case counts.

\autoref{tab:rq3-2} presents the average time to first bug detection for different bug categories. 
The first column lists the bug category, while the second column shows the average number of test cases required to detect the first instance of each bug type, and the third column shows the corresponding time in minutes.

\begin{figure}[]
  \centering
  \scalebox{0.7}{
  \includegraphics[trim={0cm 0cm 0cm 0cm},clip,width=0.95\linewidth]{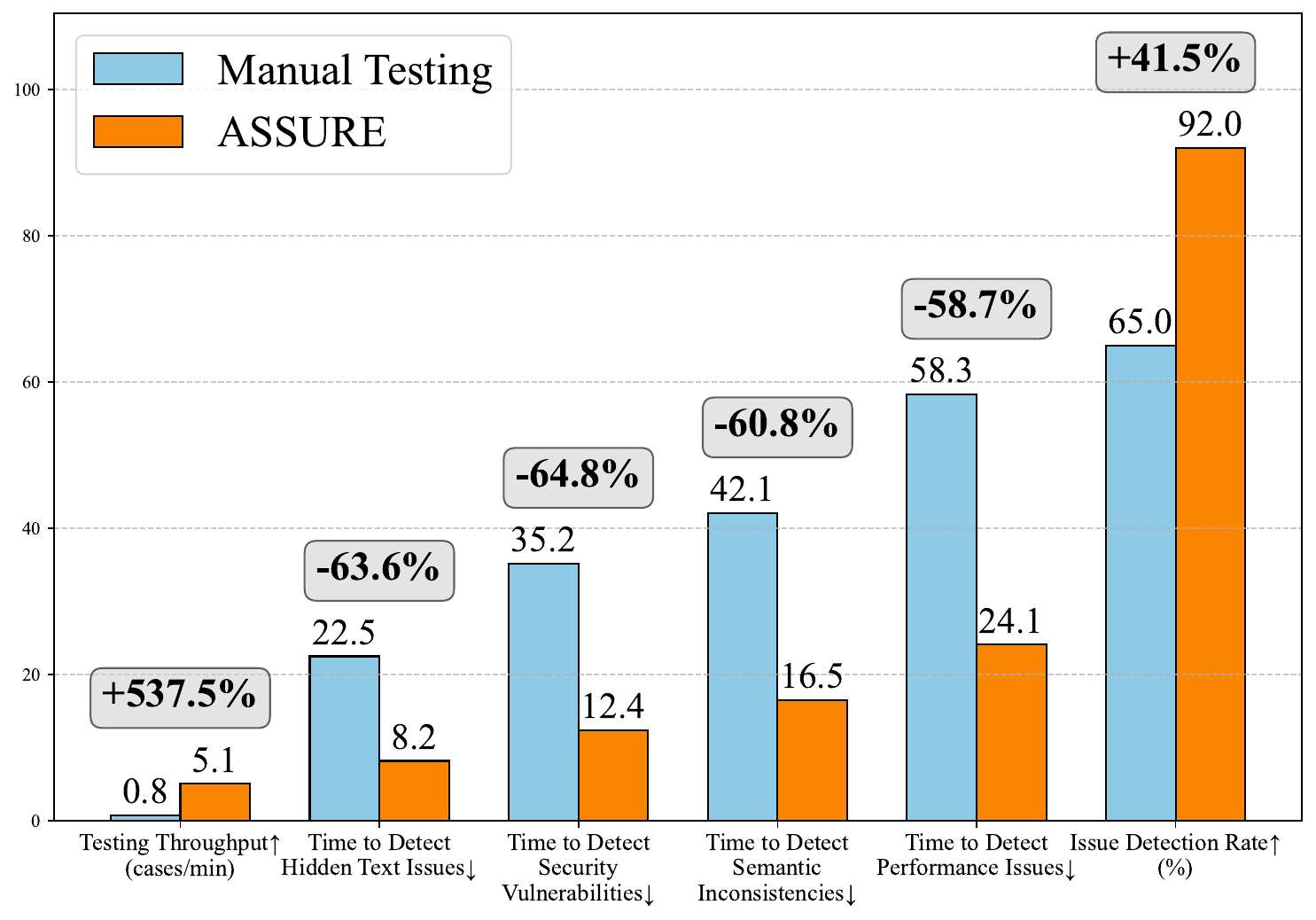}
  }
  \caption{Comparison of testing efficiency between manual testing and \sys system across six key performance metrics.}
  \label{fig:efficiency-comparison}
%   \vspace{-10pt}
\end{figure}

\begin{figure*}[]
    \centering
    \scalebox{0.7}{
    \includegraphics[trim={1cm 0cm 0cm 0cm},clip,width=0.95\linewidth]{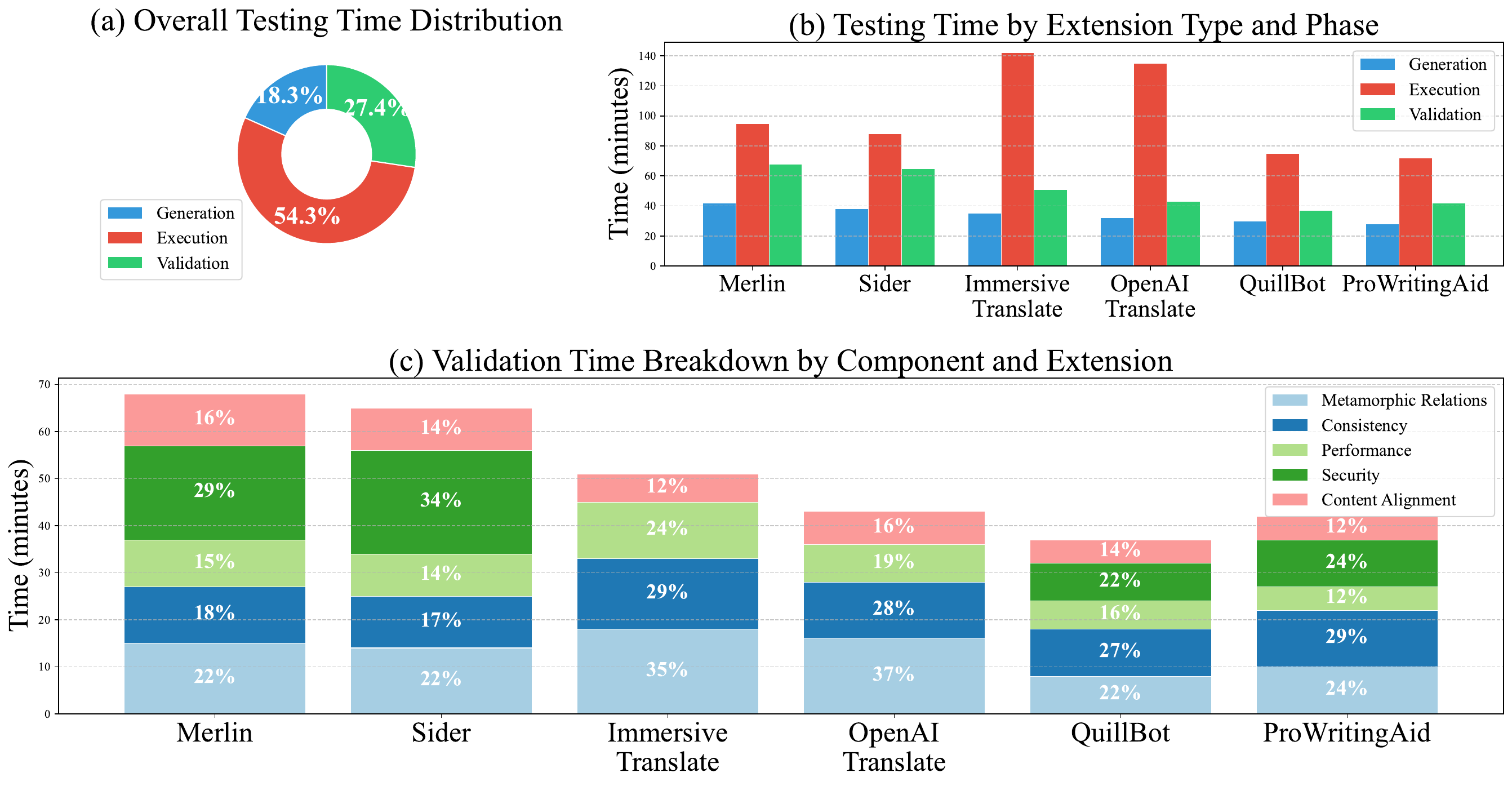}
    }
    \caption{Breakdown of \sys testing time.
    }
    \label{fig:time-breakdown}
    \vspace{-10pt}
  \end{figure*}

Figure~\ref{fig:time-breakdown} illustrates how testing time is distributed across the three main phases of the \sys system and how this distribution varies by extension type. 
As shown in the figure, test execution consumes the majority of total testing time (54.3\%), followed by validation (27.4\%) and generation (18.3\%). 
This distribution varies notably across different extension categories, with translation extensions requiring substantially more execution time than summarization or writing assistance extensions.

Figure~\ref{fig:efficiency-comparison} presents a direct comparison between manual testing and the \sys system across six key efficiency metrics. 
The horizontal axis displays six performance indicators: testing throughput~(cases/minute), time to detect various issue types~(minutes), and overall issue detection rate. 
Blue bars represent manual testing performance, while orange bars represent the \sys system. The percentage values between each pair of bars indicate the relative improvement achieved by \sys compared to manual testing. 
For time-based metrics~(middle four categories), lower values indicate better performance, while for throughput and detection rate~(first and last categories), higher values indicate better performance.

These data clearly demonstrate \sys's superior performance in all dimensions, with particularly notable advantages in testing throughput and the speed of detecting critical issues such as hidden text problems and security vulnerabilities.

\noindent
{\bf Analysis}:
The efficiency results reveal several important patterns about \sys's testing performance.

\autoref{fig:time-breakdown}(a) provides a comprehensive view of how testing time is distributed across the three main phases of the \sys system. Test execution dominates the process, consuming 54.3\% of the total testing time, while validation and generation phases account for 27.4\% and 18.3\%, respectively. 
This distribution highlights the computational intensity of the test execution phase, which involves rendering pages, executing browser extensions, and capturing their behaviors in controlled environments.
As shown in \autoref{fig:time-breakdown}(b), the time distribution varies significantly across different extension types. 
Translation extensions (Immersive Translate and OpenAI Translate) exhibit substantially longer execution times compared to summarization or writing assistance extensions. 
At our standard test volume of 1000 cases per extension, complete testing required between 142.2 minutes (ProWritingAid) and 228.4 minutes (Immersive Translate), with translation extensions taking approximately 45\% longer than writing assistance extensions. 
This difference stems primarily from the execution phase, where translation extensions process entire page content rather than selected sections, resulting in longer processing times.
The validation time breakdown in \autoref{fig:time-breakdown}(c) reveals distinct validation priorities across extension categories. Summarization extensions (Merlin and Sider) dedicate significant validation time to security testing (dark green segments), which is completely absent in translation extensions. 
Conversely, translation extensions allocate more time to metamorphic relation validation (light blue segments) and consistency checking (dark blue segments). 
\sys demonstrates linear scaling with test case count, with testing time increasing proportionally to the number of test cases. 
This predictable scaling is important for planning testing resources and schedules. 
Across all extensions, we observed an average throughput of 5.1 test cases per minute.
% Translation extensions~(Immersive Translate and OpenAI Translate) required the most testing time~(228.4 and 209.8 minutes respectively for 1000 tests), approximately 45\% longer than writing assistance extensions~(ProWritingAid at 142.2 minutes). 
% This difference stems primarily from the execution phase, where translation extensions process entire page content rather than selected sections, resulting in longer processing times and more complex output validation.

The time to first bug detection analysis reveals significant efficiency differences across bug categories. 
Hidden text processing issues were detected most quickly~(8.2 minutes on average), reflecting the effectiveness of our content alignment validation in identifying these problems with relatively few test cases. 
Performance degradation issues required substantially more time to detect~(24.1 minutes), as they often emerge only after exercising extensions with multiple complex inputs. 
Security issues, including prompt injection vulnerabilities, were detected relatively quickly~(12.4 minutes), demonstrating the effectiveness of our security validation component in identifying these critical issues.

When compared to manual testing approaches, as shown in Figure~\ref{fig:efficiency-comparison}, \sys demonstrates substantial efficiency improvements. Manual testing achieved an average throughput of only 0.8 test cases per minute, compared to \sys's 5.1 test cases per minute—a 6.4x improvement. 
The \sys system also detected critical security vulnerabilities in 12.4 minutes on average, compared to 35.2 minutes for manual testing~(a 2.8x speedup). 
Moreover, manual testing showed significantly lower detection rates for subtle issues such as inconsistent behavior across semantically equivalent inputs, with an overall issue detection rate of 65\% compared to \sys's 92\%.
These efficiency characteristics make \sys practical for integration into development pipelines and continuous integration systems, where it can provide comprehensive testing of AI-powered browser extensions without imposing prohibitive resource or time requirements.

\section{Discussion}\label{sec:disc}
Our findings through \sys reveal several critical patterns of vulnerability in AI-powered browser extensions that have significant implications for developers. 
Based on our systematic evaluation of 531 identified issues across six popular extensions, we offer the following recommendations to guide more secure and reliable AI browser extension development.

\textbf{Visible-only processing.}
Our analysis of hidden text processing vulnerabilities suggests developers should adopt a ``visible-only processing'' principle. 
This pattern ensures that AI components only process content visible to users, aligning extension behavior with user expectations. 
Implementing DOM visibility filters before passing content to AI models can prevent the inadvertent processing of hidden elements that could lead to misleading or manipulated outputs. 
For example, content summarization extensions should employ explicit visibility checks using computed styles and positioning attributes to exclude hidden content from consideration. 

\textbf{Input sanitization.}
The prompt injection vulnerabilities detected across multiple extension categories highlight the need for robust input sanitization techniques. 
Developers should implement multi-layer defense strategies, such as pattern-based filtering to detect and neutralize potential injection commands, contextual boundaries in prompts sent to underlying models, and output verification to detect unexpected instruction-following behavior.

\textbf{Consistency enforcement.}
The inconsistent behavior observed across semantically equivalent inputs calls for semantic equivalence checks in browser extensions. 
Developers should implement transformer-based semantic similarity measures to ensure stable processing regardless of DOM structure variations.

\textbf{Loading strategies Optimization.}
The performance degradation patterns detected in our evaluation point to the importance of implementing chunking and progressive loading strategies for large content processing. 
Extensions should adopt adaptive processing approaches, such as breaking large content into manageable segments, processing content progressively as users interact with the page, and applying caching strategies for previously processed content.
Our performance validation results indicate that extensions employing chunking strategies demonstrated linear scaling patterns rather than the exponential degradation observed in monolithic processing approaches. 
This approach is particularly critical for translation extensions handling large documents.

\section{Threat to Validity}\label{sec:threat}

\sys is evaluated on six popular AI-powered browser extensions across three categories~(content summarization, language translation, and writing assistance), which may present limitations in scope. 
Additionally, the presence of configurable validation components and test generation parameters introduces some threats to the generalizability of our results.
While our experiments demonstrated promising results in detecting various bug categories, the effectiveness remains uncertain when applied to extensions built with emerging AI models or those serving specialized domains not covered in our evaluation. 
The dynamic nature of web content and the rapid evolution of LLM capabilities further challenge the long-term validity of specific test patterns.

Our evaluation also revealed limitations regarding false positives and detection capabilities. 
We observed false positives primarily in metamorphic relations~(8.3\% false positive rate) and content alignment~(12.7\% false positive rate), often stemming from the inherent variability in AI outputs or extensions' intentional processing of hidden metadata. 
We mitigated these issues through adaptive thresholds and refined validators, though challenges remain. 
Detection limitations include difficulty identifying subtle biases in extension outputs, limited evaluation of extensions handling highly dynamic JavaScript-rendered content, and reduced effectiveness against sophisticated prompt injection attacks employing subtle semantic manipulations rather than direct instruction patterns.

To address these concerns, we have open-sourced the implementation of \sys and provided comprehensive experimental details including test case generation templates, validation configurations, and execution parameters. 
Our modular design allows for extension to accommodate new bug patterns and extension categories as they emerge. 
For reproduction purposes, all code, configurations, and collected datasets are available at~\cite{AnonymizedRepositoryAnonymousc}.

\section{Related Work}\label{sec:rw}

\subsection{Metamorphic Testing}\label{sec:metamorphic}

Metamorphic Testing~(MT) has become a mainstream approach to tackle the test oracle problem, i.e. situations where no reliable expected-output oracle exists~\cite{xuMetamorphicTestingNamed2022}.
Chen et al. ~\cite{chenMetamorphicTestingNew2020} pioneered this approach in the late 1990s, introducing a systematic methodology for testing programs where expected outputs are difficult to verify. 
In the software engineering domain, Zhou et al. ~\cite{zhouMetamorphicTestingSoftware2016} applied metamorphic testing to search engines, demonstrating its effectiveness in quality assessment of complex web applications. 
Segura et al. ~\cite{seguraMetamorphicTestingRESTful2018} extended this work to RESTful web APIs by developing metamorphic relation output patterns. 
For concurrent programs, Sun et al. ~\cite{sunInterleavingGuidedMetamorphic2023} introduced an interleaving-guided metamorphic testing approach using specialized patterns to detect concurrency bugs.
The category-choice framework proposed by Chen et al. ~\cite{chenMETRICMETamorphicRelation2016} with their METRIC technique provided a systematic approach to construct metamorphic relations by comparing different test scenarios. 
Sun et al. ~\cite{sunMETRICMetamorphicRelation2021} later enhanced this with METRIC+, incorporating both input and output domains for more effective relation identification. 
For cybersecurity testing, Chaleshtari et al. ~\cite{chaleshtariMetamorphicTestingWeb2023} developed a specialized language to express security-focused metamorphic relations.
Tian et al. ~\cite{tianDeepTestAutomatedTesting2018a} used metamorphic testing to verify deep neural network-driven vehicles, while more recently, Deng et al. ~\cite{dengBMTBehaviorDriven2021,dengDeclarativeMetamorphicTesting2023} developed declarative behavior-driven approaches for autonomous systems testing.
The integration of large language models represents the newest frontier, with Zhang et al. ~\cite{zhangAutomatedMetamorphicRelationGeneration2023} leveraging ChatGPT to automatically generate metamorphic relations for autonomous driving, demonstrating both the potential and limitations of AI-assisted relation generation. 
Liu et al. ~\cite{liuGenerationbasedDifferentialFuzzing2023} focused specifically on deep learning libraries, developing differential fuzzing techniques based on symmetry metamorphic relation patterns.

\subsection{Browser Extension Testing}\label{sec:BETesting}

Browser extensions have elevated privileges that make them an attractive target for attackers, leading to significant research on their security. 
Bandhakavi et al.\cite{bandhakaviVettingBrowserExtensions2011} introduced Vex, the first static information flow tracking tool for 2,452 Firefox XPCOM extensions, though it predated modern message-passing APIs. 
Carlini et al.\cite{carliniEvaluationGoogleChrome2012} combined network traffic analysis with manual review of 100 Chrome extensions to evaluate Chrome's security mechanisms. 
Calzavara et al.~\cite{calzavaraFineGrainedDetectionPrivilege2015} proposed a formal security analysis framework to assess privilege escalation risks if specific extension components were compromised.
For detecting vulnerabilities in browser extensions, Buyukkayhan et al.\cite{buyukkayhanCrossFireAnalysisFirefox2016} developed CrossFire, a static data flow analysis tool to identify dangerous flows between globally accessible extension variables and security-sensitive XPCOM calls. 
Starov et al.\cite{starovExtendedTrackingPowers2017} conducted a dynamic analysis with BrowsingFog on 10,000 Chrome extensions, revealing that most privacy leakage was unintentional rather than malicious. 
Researchers have also focused on detecting malicious extensions through behavior monitoring~\cite{kapravelosHulkElicitingMalicious2014}, anomalous rating detection~\cite{pantelaiosYouveChangedDetecting2020}, developer reputation tracking~\cite{jagpalTrendsLessonsThree2015}, user tracking~\cite{chenMystiqueUncoveringInformation2018}, and privacy violation detection~\cite{chenMystiqueUncoveringInformation2018}.

\subsection{LLMs Testing}\label{sec:LLMsTesting}

Large language models~(LLMs) have become increasingly important in various applications, leading to significant research on their security and testing methodologies. 
For detecting and evaluating vulnerabilities in LLMs, Deng et al.\cite{dengMasterKeyAutomatedJailbreak2024} introduced MasterKey, an automated methodology for creating jailbreak prompts that systematically attack commercial LLM chatbots. 
Liu et al.\cite{liuAutoDANGeneratingStealthy2024} developed AutoDAN, which generates semantically meaningful jailbreak prompts against aligned LLMs and demonstrates robustness against perplexity-based defense techniques. 
Liu et al.\cite{liuPromptInjectionAttack2024} conducted pioneering work on prompt injection attacks against LLM-integrated applications, demonstrating how malicious users can bypass safety measures through carefully crafted prompts. 
CheckList~\cite{ribeiroAccuracyBehavioralTesting2020} moved beyond accuracy metrics by introducing behavioral testing for NLP models with task-specific templates, revealing vulnerabilities even in state-of-the-art systems.
For detecting limitations in LLMs' robustness, PromptBench~\cite{zhuPromptRobustEvaluatingRobustness2024} provided a comprehensive framework for evaluating adversarial prompt sensitivity at multiple levels~(character, word, sentence, and semantic). 
Wang et al.\cite{wangRobustnessChatGPTAdversarial2023} conducted an early assessment of ChatGPT's vulnerabilities to both adversarial and out-of-distribution inputs. 
GLUE-X\cite{yangGLUEXEvaluatingNatural2023} and BOSS~\cite{yuanRevisitingOutofdistributionRobustness2023} specifically focused on evaluating out-of-distribution robustness, revealing limitations when models face distribution shifts.
Researchers have also developed specialized evaluation frameworks for LLMs. 
HELM~\cite{liangHolisticEvaluationLanguage2023} provided a holistic assessment across dimensions like accuracy, robustness, calibration, and fairness. 
Big-Bench~\cite{srivastavaImitationGameQuantifying2023} introduced 204 challenging tasks to quantify capabilities beyond standard benchmarks. 
PandaLM~\cite{wangPandaLMAutomaticEvaluation2024} enabled automated language model assessment through a discriminative LLM specifically trained to evaluate multiple high-proficiency models with consideration for both objective correctness and subjective elements like clarity and comprehensiveness.

\section{Conclusion}\label{s:conclusion}

We have presented \sys, a novel metamorphic testing framework designed for AI-powered browser extensions. 
\sys bridges the critical gap between traditional extension testing and AI system validation through its modular architecture of test case generation, configurable validation, and automated execution. 
Our evaluation demonstrates \sys's effectiveness in detecting critical issues including hidden text manipulation, prompt injection vulnerabilities, inconsistent behavior, and performance degradation across various extension types. 
By providing a systematic approach to testing the complex interactions between AI models and browser environments, \sys improves the reliability and security of AI-powered extensions while establishing a foundation for more robust extension development in the future.

\section{Data Availability}\label{s:data}
% To follow the Open Science Policy and support reproducibility, we have released code about our implementations and evaluations.
All source code and data used in our work can be found at~\cite{AnonymizedRepositoryAnonymousc}.

% \section*{Acknowledgement}\label{s:ack}

% We thank the anonymous reviewers for their constructive comments. 
% This research is supported by the National Key Research and Development Program of China (2023YFB3107400), the National Natural Science Foundation of China (62376210, 62161160337, 62132011, U21B2018, U20A20177, 62206217), the Shaanxi Province Key Industry Innovation Program (2023-ZDLGY-38, 2021ZDLGY01-02).
% Chao Shen is the corresponding author. 

\newpage
\bibliographystyle{ACM-Reference-Format}
\bibliography{REFLIST}

\end{document}